\documentclass[12pt,preprint]{aastex}

\shorttitle{X-ray observations of CP Pup}
\shortauthors{Orio et al.}

\def\mdot{$\dot m$}
\begin{document}


\title{New X-ray observations of the old nova CP Puppis and of the more
 recent nova V351 Pup}


\author{M. Orio\altaffilmark{1,2}}
\affil{INAF - Osservatorio Astronomico di Padova,
    vicolo Osservatorio, 5, I-35122 Padova, Italy}
\email{orio@astro.wisc.edu}

\author{K. Mukai \altaffilmark{3}}
\affil{CRESST and X-ray Astrophysics Laboratory NASA/GSFC, Greenbelt, MD 20771, USA}

\author{A. Bianchini}
\affil{Department of Astronomy, Padova University, vicolo Osservatorio, 3,
  I-35122 Padova, Italy }

\author{D. de Martino}
\affil{INAF - Osservatorio Astronomico di Capodimonte, via Moiarello 16,
 I-80131 Napoli, Italy}

\and

\author{S. Howell}
\affil{NOAO, 950 N. Cherry Ave., Tucson, AZ, USA}

\altaffiltext{1}{
and Department of Astronomy, University of Wisconsin, 475 N. Charter Str., Madison WI 53706}
\altaffiltext{2}{ visitor at the Kavli Institute for Theoretical Physics, Kohn Hall, Santa
 Barbara, CA 93106}
\altaffiltext{3}{and Department of Physics, University of Maryland,
 Baltimore County, 1000 Hilltop Circle, Baltimore, MD 21250, USA}

\begin{abstract}
 We present X-ray observations of the field 
containing Nova Puppis 1942 (CP Pup) and  Nova Puppis 1991 (V351 Pup),
 done with {\sl ASCA} in 1998, and with {\sl XMM-Newton}  in 2005. 
 The X-ray and UV luminosity of CP Pup seem to have remained 
 approximately constant since  the last X-ray observations of the 1980'ies, while 
 the optical luminosity has decreased.
The X-ray properties of this nova are explained by a 
high mass white dwarf accreting at low rate, in agreement with the nova theory given 
 the large amplitude and other characteristics
 of the 1942 outburst. 
 Assuming a distance of 1600 pc, the X-ray luminosity
 of CP Pup is L$_{\rm x}=$2.2 $\times 10^{33}$ erg s$^{-1}$ in the
 0.15-10 keV range covered with EPIC, compatible with a 
 magnetic system. The RGS grating spectrum shows 
  a few prominent emission lines, and
it is fitted with  a cooling flow with mass accretion rate 
 \mdot$< 1.6 \times 10^{-10}$ M$_\odot$ yr$^{-1}$. We detected
also the O VII complex at 21.6-21.8 \AA \ that does not arise in the 
cooling flow. Most likely this feature originates in a wind
or in the nova shell.
 The RGS and EPIC  spectra are fitted only with thermal models with  
 a very high shock temperature, T$>$60 keV, 
indicating  a white dwarf with M$>$1.1 M$_\odot$.
 The X-ray flux is modulated with the spectroscopic period of 1.47 hours detected
 in the optical. Since CP Pup is not an eclipsing system, 
 this is better understood if magnetic accretion occurs: 
 we discuss this possibility and its implications in detail.
V351 Pup (N Pup 1991) was detected  with 
 {\sl XMM-Newton}, but not with {\sl ASCA}. It is  
 a faint, non-super-soft X-ray source with luminosity
L$_{\rm x} \simeq 3 \times 10^{31}$ erg s$^{-1}$,
 a factor of 50 less than measured with ROSAT in 1993.  
\end{abstract}
\section{Introduction}
 In this article we describe {\sl ASCA} and {\sl XMM-Newton} observations
 of two novae in the same field in the Puppis region,
 Nova Puppis 1942 (CP Pup) and Nova Puppis 1991 (V351 Pup).
 Classical novae are Cataclysmic Variables (CV), that is
 close, interacting
 binaries containing a white dwarf (WD) that accretes matter from
 a companion, which is usually still on the main sequence or 
 slightly evolved. In novae, at the bottom of the accreted envelope
hydrogen burning is ignited in a thin shell, first the p-p and later 
 the CNO cycle. The process becomes explosive because of
 the degeneracy conditions of the material, compressed on the small
 surface of the WD.  A thermonuclear flash follows, and an optically thick wind that 
 ejects most or all the accreted material (see Starrfield
 et al. 2000 for a review).
 The more massive the WD is, the smaller its surface,  and the 
 sooner the pressure sufficient to initiate the thermonuclear runaway
 is reached.  The nova theory makes very
 definite predictions on the correlation of the binary system
 parameters and the properties of the outburst. The main parameters
 that determine its properties and recurrence time 
 are the mass accretion rate \mdot \
 and the white dwarf mass m(WD) (note that all novae
 are recurrent phenomena even if only very few, the ``recurrent novae''
 are repeated over a  human lifetime). The abundances and the thermal state
 of the WD at the onset of accretion are also important.
   In a series of papers (especially Kovetz \& Prialnik 1994, Prialnik \&
 Kovetz 1995, Yaron et al. 2004, Epelstein et al. 2007) clear 
 correlations between these parameters and the outburst characteristics
 are found. Thus, having a remarkable and detectable manifestation in 
 their outburst, novae allow us to test the evolutionary
 models that apply to all CV 
  and to related systems, included the  ``single degenerate''
 progenitors of type Ia supernovae. 
  
 X-ray observations are a powerful tool to study 
 the following phenomena in novae: a) Violent, energetic mechanisms in the
 shell shortly after the outburst
 (e.g. Nelson et al. 2008), or even a long time later  (Balman 2006);
 b) The WD atmospheric temperature, abundances and effective
 gravity, after the ejecta clear up sufficiently to
 allow detection of the X-ray supersoft, luminous central source (e.g.
 Ness et al. 2003, Nelson et al. 2008); c) The accretion
 process in quiescence. The shocked material in the
 disk, or in a magnetically funneled accretion stream, emits X-rays
 (e.g. Mukai \& Orio 2005). In order to study accretion,
 we proposed X-ray observations of the two objects
 we describe in this paper, as a test of the nova theory which
 is largely based on the modalities of accretion.

\subsection{CP Pup: a puzzling nova}
 CP Puppis appeared as a truly astonishing object in 1942,
 leaving the astronomers of that  time wondering for a while whether it 
 was a Galactic supernova. The
outburst is described in the Payne-Gaposhkin's
 book ``Classical Novae''. In
 1942 it rose from fainter than 17th magnitude to  V=-0.2,
 raising to maximum during at least 3 days after the discovery. The 
time for a decay by 3 magnitudes (t$_3$) was one of the shortest
 ever measured for a nova,
 only 6.5 days. The velocities measured from the full width at
 half maximum of 
 the initial absorption lines, and of the emission lines that
 appeared soon thereafter, reached at most 1200 km s$^{-1}$, which 
 is lower than measured for other luminous novae. 
 The ejected shell was resolved for the first time 14 years after the outburst
(see Williams 1982 and references therein),
 indicating a distance of about 1.6 kpc and a maximum absolute luminosity
  M$_{\rm V}$=-10.5, a factor of $\simeq$400 above the Eddington luminosity
 of a 1 M$_\odot$ star (L$_{\rm Edd}$).  Downes \&
 D\"urbeck (2000) discussed the possible
 errors in determining the  distance of a shell that
 is not expanding uniformly, but is made of
 blobs with different velocity. These authors placed
 CP Pup at a lower distance of 1140 pc, and Cohen \& Rosenthal (1983) even 
 at only 850 pc. A distance compatible with a peak luminosity 
 below 100 L$_{\rm Edd}$ is in any case very unlikely. Another unusual
 characteristic of CP Pup were the Fe [II] lines that still
 appeared in the   spectrum  simultaneously with high excitation lines,
 including even coronal lines. The energetics initially reminded of
a supernova. 
If we were witnessing such an event today, probably we
 would initially suspect an accretion induced collapse
 of a white dwarf, or merging compact objects. 
 Despite the peculiarities outlined above, 
 the optical spectrum was typical for a classical nova,
and as such CP Pup is classified. 

The distance obtained from the  maximum magnitude versus
 rate of decline relationship (MMRD) 
is  much larger than the range of
 values derived from the nebular parallax.
For this and several other reasons CP Pup has long been
 suspected to host a  magnetic WD. Because of the apparent disk-like
 structure, it may be an 
 ``intermediate polar'' (hereafter IP) rather than a ``polar'' (e.g. 
 a diskless system, see
discussion in Balman et al. 1995 and references therein).  
It is doubtful whether the MMRD holds for magnetic novae (see
for instance the description of DQ Her by Payne Gaposhkin, 1964).
Orio et al. (1992) suggested that magnetic rotator effects
 may cause the ejection of
 the accreted envelope more quickly and efficiently than in other
 novae.
In fact the outburst amplitude of V1500 Cyg, a 
 polar, was also very large, about 16 mag, and  the outburst of
the IP GK Per  had an amplitude of $\simeq$14 mag despite the 
 high luminosity at quiescence (due to an unusual subgiant secondary).
 The  unambiguous classification of also CP Pup as 
 a magnetic system would be further evidence of the 
 role of the magnetic field in nova outbursts.

 Several unusual characteristics of the quiescent CP Pup point in the direction 
of an IP nature.
 The outburst of 1942 can best be explained 
 with a high mass white dwarf (e.g., Prialnik and Kovetz 1995,
 see discussion below).  However, assuming that the
 radial velocities variations of the emission lines in the quiescent
optical spectrum indicate the orbital motion 
 of the compact object, and that the width of the lines is due to the 
Keplerian velocity of the disk,  an
 embarrassingly low mass of the white dwarf is obtained, M$<$0.2 M$_\odot$.
  The data are rather complex, and
 we will summarise them as follows. At quiescence,
 two photometric periods have been measured, 
 0.061-0.062 days, and 0.06834 days (Bianchini
 1985a and b, Warner 1985, O'Donoghue et al. 1985, White et 
al. 1993). 
  From the radial velocity
of the He II line at 4686 \AA \ and the Balmer
 lines, White et al. (1993) measured a spectroscopic period of 0.06129
 days.  In the most recent work, Bianchini et al.
 (2008, in preparation, and hereafter B08) detect a 
 mean spectroscopic period of 0.0612643 days (1.47 hours),
 coincident within the errors with
 the period reported by White et al. (1993), and other aliases of this period.
 In 1987 a photometric period of CP Pup was measured to be
11\% longer than the spectroscopic period (White et al. 1993). However 
 in 1993 Patterson \& Warner (1998) found instead two other
 photometric periods, of which the first was
 equal to the spectroscopic one 
 and the second was unstable, but generally  longer than the
 spectroscopic period by only 
 $\simeq$2\%. The last authors interpret this as
 a ``superhump'' (due to disk precession), which is often detected in
non-magnetic cataclysmic variables (CV).
However, another possible interpretation is that the
longer photometric period is due to the rotation of the WD
that has become asynchronous after the nova outburst,
like the polar Nova V1500 Cyg after the 1975 outburst (Diaz \& Steiner 1991).
Balman et al. (1995) reported an  X-ray flux modulation with
the spectroscopic period. This is suggestive of a
magnetic system, since X-ray orbital modulations are not
 observed in non-magnetic
CV with inclination less than 60$^{\rm o}$ (Baskill et al. 2005).
The evidence however was not conclusive, because the orbital
 period of CP Pup is uncomfortably close to the rotation period
of 94 minutes of the {\sl ROSAT} spacecraft.

The spectroscopic period of 1.47 hours is interpreted as the orbital
period in the literature, which we also adopt as our baseline
interpretation (we will review this assumption later, and present
an alternative interpretation).  In this baseline scenario, it is
one of the two shortest orbital periods of all novae. 
Most of the known orbital periods of classical novae are 
longer than 3 hours, i.e. above the period gap.
 CP Pup is only matched by GQ Mus (Nova Mus 1983)
 with 1.42 hours, and there are only 5 other novae
 with orbital periods below 3 hours (Diaz \& Bruch 1997).

The lack of eclipses places a constraint on the inclination, 
$i\leq$65$^{\rm o}$.
 With the measured radial velocities and
line widths, there simply is no way to place a massive
 white dwarf in a non-eclipsing system with such
 a short orbital period. O'Donoghue et al. (1989)
 conclude that the WD mass should be M$_{\rm WD}\leq$0.2 M$_\odot$.
 Moreover, if the modulation of the infrared flux with
 a period close to the spectroscopic one is due to an ellipsoidal
 variation (Szkody \& Feinswog 1988), the upper limit  
 on the inclination is {\it i}$\leq$35$^{\rm o}$.
 This implies that a Roche lobe filling secondary secondary 
has mass $\leq$0.2 M$_\odot$, and that the upper limit  
 for the WD is less than a tenth of a solar mass!
This is  a clear inconsistency:
in the first place, this WD mass is much lower than the minimum WD mass
in the Galaxy and it is difficult to explain such an amount
of mass loss from the WD. Moreover,
H-burning would not even start on the surface of such a low
 mass  WD, thus, a classical nova eruption could not occur.
This puzzle may originate, first of all, in an incorrect interpretation of the 
 IR modulation.
 At a distance of $\sim$1 kpc, a 0.2 M$_\odot$ secondary star (M5V,
M$_{\rm V}$=14.7, M$_{\rm K}$=8.5) would have a distance modulus
 m-M=10, yielding  
a K$\simeq$18.5 magnitude of the secondary. Szkody and Feinswog measured
K=13.4 for CP Pup, a value completely inconsistent with the assumed
short orbital period secondary and one which cannot be reconciled as
ellipsoidal variations. Only if CP Pup was 95 pc away, the assumed
0.2 M$_\odot$ secondary star could then account for the observed IR
modulations.  A different physical mechanism
 that may cause the IR light curve modulation is beamed cyclotron radiation.
Although up to now the cyclotron humps in IR have 
 only been observed for the higher magnetic field systems, the polars,
 this phenomenon is not yet ruled out in  IP's. The 
 cyclotron hump does produce an IR modulation like the one
 observed in CP Pup. One example is  AR UMa (Howell et al. 2001),
 where the visual and infrared double-humped light curves
 are caused by beamed cyclotron radiation, although initially at least the
 visual modulations
 were initially ascribed to ellipsoidal variations.
 Another example is HU Aqr, another polar, where the IR light
 curve is explained partially by cyclotron emission and
 the   light curve is very complex to model (Howell et al. 2002).

The secondary star spectrum is not detected in optical or
 infrared, therefore the secondary is  thought to be a main sequence star,
because subgiants or giants have clear spectral signatures at distances of
 order of 1 kpc.
If the measured spectroscopic period is 
 orbital and CP Pup is not a triple or multiple
 system, the most likely reason for the absurd result of the
 disk accretor model is simply that the
 accretion is funneled by a magnetic field and does not
 occur through the disk. The saddled line profiles 
 show that a disk exists, however
if the innermost part of the accretion disk
is disrupted by the magnetic field of an IP, the radial velocity 
 is not measured close to the WD, but at the Alfv\`en radius.

Finally, we do not rule out that the radial velocities measurements are 
 inaccurate because the emission
lines are contaminated by multiple components.
 The optical spectrum does show very complex velocity profiles,
 with multiple components in the emission lines (B08);
some of these components may be produced in a hot spot on the WD
or in  an accretion stream rather than in the disk.
 B08 describe these possibilities in detail,
but they conclude that assuming that CP Puppis is an 
 IP and that the 1.47 hours period is orbital 
does not completely resolve the issue.  
 Analysing new spectroscopic and photometric data taken
 over several years, B08 find that the structure of the emission
lines is very complex and the velocity of the inner disk was not
 accurately determined, 
 alleviating somewhat the ``mass function problem'', but at the same
 time making the study of the system extremely difficult.

When we  proposed observing CP Pup 
 with {\sl XMM-Newton}, one of our aims was to resolve the mass problem,
 possibly revealing the spin period of a magnetic WD.
 One of the characteristics signatures of IP systems is in fact
 that the X-ray flux is usually modulated with the rotation
 period of the WD, which does not rotate synchronously with
 the orbital period like in ``polar'' systems. Typical
 spin periods of IP are of the order of tens of minutes.
  A new X-ray observation of CP Pup also had a broader scope
 than learning the details of a single system.
 Because of the selection effect due to the outburst amplitude, classical
 novae are observed at larger distances than other CV, and
 they are rather faint at quiescence. Up to now, only one 
classical nova, V603 Aql, has yielded a grating spectrum in X-rays
(Mukai \& Orio 2005). X-ray grating spectra are a fundamental tool
 in order to study how accretion occurs. Therefore, another important 
 aim we had was to broaden the statistics, obtaining another
 term of comparison to learn how accretion occurs in novae.  CP Pup
 is the third brightest classical nova in X-rays at quiescence,
 after V603 Aql and GK Per. 

\subsection{V351 Pup: a cooling nebula?}

V351 Pup was discovered in outburst 
in 1991 December 27  (Camilleri 1992). It was a moderately fast classical nova,
 with t$_2$ and t$_3$ of 16 and 40 
days, respectively and ejection velocity $\simeq$2000 km s$^{-1}$ on average
(Shore et al. 1992, Sonneborn et al. 1992).
 It was also a neon nova (Pachoulakis \& Saizar 1995). 

An X-ray flux in the 0.2-2.4 keV range F$_{\rm x}=3 \times 10^{-12}$ erg
cm$^{-2}$ s$^{-1}$ was 
 measured with {\sl ROSAT} 16 months after the outburst (Orio et al. 1996).
The spectrum was quite hard in the {\sl ROSAT}
 range and Orio et al. (1996) suggested
 that the X-ray emission could originate either in 
the cooling nova shell, or be due to  renewed accretion. In Orio
 et al. (2001) we already announced and discussed that the nova
 was not detected
with {\sl ASCA}, with a strict upper limit of 10$^{-13}$ erg cm$^{2}$
 s$^{-1}$ in the 0.7-10 keV
 energy range. We concluded in that paper that the first interpretation
 is definitely the most likely, because the shell cools and disappears as X-ray
 source, while accretion is expected to continue.
 We present here a new detection of V351Pup with XMM-Newton, but the 
 data are far poorer in quality than the CP Pup ones, so 
 we will discuss this object in less detail.

\section{The new X-ray observations of CP Pup}

 In the first rows of Table 1 (before the double line dividing
 the Table in two parts) we give details  
 concerning the X-ray observations of CP Pup done before 1998. 
 The new observations are below the double line.
   The first of these observations was serendipitously made with the {\sl ASCA} GIS
(Gas Imaging Spectrometers, see Tanaka et al.
1994) on 1998/11/6, during a  
50 ksec long exposure of V351 Pup. CP Puppis was in the GIS 
 field of view.
The second  new observation, of
 much better quality, was a pointing of CP Pup with {\sl XMM-Newton}
on 2005/6/4, with an exposure time slightly longer than 50 ksec. 
V351 Pup was at the very edge of the field of view of the EPIC-pn and
 MOS-1.
The {\sl XMM-Newton} data are exceptionally good for
 a relatively faint CV like CP Pup. A grating
 spectrum, albeit with low S/N, has been observed with the RGS. 
 CP Pup is only the second quiescent nova for which a
grating X-ray spectrum exists at quiescence, although the S/N is
 not as high as it was for V603 Aql (Mukai \& Orio 2005).

\subsection{The {\sl ASCA} observation}

 Since CP Pup was  15 arcminutes off-axis, it was
 observed only with the {\sl ASCA} GIS detectors, but  not
 with the SIS. The data reduction was performed with FTOOLS,
the spectra were fitted with XSPEC (Arnaud 1996). The spectra
 are shown in Fig. 1. The quality of the data is poor, but it allows
 a useful comparison with the previous {\sl Einstein}
 and {\sl ROSAT} observations (see first part of Table 1). 
Table 1 gives the count rates
and the unabsorbed flux derived assuming the best-fit
single components thermal models of Balman et al. (1995) for the  {\sl ROSAT}
data.  The best fit with three temperature components
 to both GIS
 detectors (with reduced $\chi^2$=1) indicates a 0.7-10 keV unabsorbed flux 
 6.2 $\times 10^{-12}$ erg s$^{-1}$.
 In the 0.7-2.4 keV band the flux is 2 $\times 10^{-12}$ erg s$^{-1}$,
within the 3$\sigma$ errors of the fits to the {\sl ROSAT} data. 
 However, with all models the fit temperature 
is          unbound, because the best fit is obtained with
the plasma temperature, or the highest temperature
 component, equal to  the upper limit allowed by
 the model. The 2$\sigma$ error in the flux indicated in the Table
 is also limited at this value.  The ASCA flux is 
 in agreement with the unabsorbed flux measured in the {\sl XMM-Newton}
 observations described below. We note that 
 three-temperature APEC model that best fits the 
 the {\sl XMM-Newton} EPIC spectra (see Table 1 and Section 2.2), 
 yields
 approximately the same parameters for {\sl ASCA}
 and {\sl XMM-Newton},  including the unabsorbed flux.
The latter has about the same value in the multiple temperature
 fit and   with
 the simple one-component models used for the initial comparison
 with {\sl ROSAT} (which is also a good fit given the quality of
 the data). 

\subsection{{\sl XMM-Newton}: The spectra}
Table 1 gives also the basic results of the {\sl XMM-Newton} observation
of CP Pup. 
 The data were extracted using XMM-SAS version 6.5 including
only single and double events (PATTERN$\leq$4), and the conservative
 screening criterion FLAG=0. 
  The spectra were fitted with XSPEC.
The RGS spectra are shown in Fig. 2, and at first glance appear
 strikingly similar to the {\sl Chandra} grating spectra of V603 Aql
 (Mukai \& Orio 2005).
 We identify a few emission lines beyond
 reasonable doubts, even if the S/N is too poor to measure
 line width and possible shifts. 
 The prominent emission lines are marked in Fig. 2:
   Ne X $\lambda$12.13, Fe XXV $\lambda$15.01, O VIII
 $\lambda$18.97 and the O VII complex at 21.5-22 \AA.
Lines of Si, possibly Si XIV $\lambda$6.18, Si XIII $\lambda$6.65,
 and Mg XII $\lambda$8.42, all observed in the V603 Aql
 spectrum (Mukai \& Orio 2005) may also be present but
 are not clearly resolved. 
 The RGS-1 and RGS-2 count rate differ
 because of the different missing chips. The RGS-2 includes the 
$\approx$10.5-14 \AA \ region where the continuum is high and there is 
the strongest line, Ne X at 12.13 \AA,  so the
count rate is higher than for the RGS-1.
 The detection of emission lines is  very important, 
 because it allows us to test whether
 the spectrum can be fitted  with  a more realistic model
 than a thermal plasma with one or more temperature components,
 a ``cooling flow''
 model. We used the MKCFLOW model in XSPEC, a steady state,
 continuously cooling
 multi-temperature plasma (see Mushotzky \& Szymkowiak 1988, 
 and Mukai et al. 2003). This model
 includes less free parameters than  multi temperature
thermal bremsstrahlung models in XSPEC,
 because it uses the emission lines strengths and ratios. 
 The parameters are the lowest and highest temperature, 
the global metal abundance, the redshift, and the mass accretion rate
through the cooling flow.  We specify a distance of 1600
pc using the redshift parameter, which the MKCFLOW model interprets
as the cosmological redshift and hence an indicator of the distance.
 MKCFLOW fits well the spectra of different types of CV, and
 especially those of non-magnetic accretors (Mukai et al. 2003).
 This model is particularly interesting to us also because,
 assuming  that all accreting matter cools via optically thin thermal
 X-ray emission, the mass accretion rate trough the 
cooling flow represents the mass transfer rate
\mdot  \ onto the WD,  a fundamental parameter in the nova thermonuclear
 runaway models. MKCFLOW yields the best fit
 to the RGS data. It fits the RGS better than the  model used
 for the EPIC data, the multi-temperature 
 APEC model in XSPEC (see description below). 
 In  Fig. 2 we how the fit to the RGS data with
 the MKCFLOW model and a single absorber, and 
 the fit parameters are given in Table 1.  This fit  is formally acceptable,
with $\chi ^2$=1, and abundances twice solar, even if the emission
 lines still seem more prominent than in the model, especially 
 Ne X at 12.13 \AA, Ne IX at 13.5 \AA, O VIII at 18.97 \AA \ and
 the He-like O VII complex at 21.5-22 \AA. The latter is typical
 of those magnetic systems whose X-ray spectrum cannot be fitted with
 the  cooling
 flow model, but it has been observed also in a non magnetic CV,
 the dwarf nova WX Hydri in which
 it has been attributed to a wind (Perna et al. 2003).
 Although with our S/N we do not measure a broadening of the O VII line
 like Perna et al. (2003) did, we note that we cannot reproduce the
observed ratio of O VIII line at 18.97 \AA \  and O VII by
 varying the temperatures and O abundance. 
We suggest that also for CP Pup
O VII is most likely originated in a wind, 
 or in the circumstellar shell of the
 ejecta. It certainly does not appear to originate in the
 accreted material.
The other lines mentioned above, especially Ne X are all 
 very prominent also in the V603 Aql spectrum (Mukai \& Orio 2005).
 For V603 Aql, Mukai \& Orio hypothesized that an overabundance of
 Ne in the material accreted from the secondary  explains the high
 flux of the Ne lines. Since CP Pup had a long-lasting extended shell
 that was repeatedly observed in optical images (e.g. Williams
 et al. 1982), we  do not rule out that the X-ray flux and the emission
 lines may be partially  
 produced also in the ejecta. The nebula of Nova GK Per
 emits X-rays even more than a century after the outburst
(Balman 2005). DQ Her and RR Pic also showed some extended X-ray
 emission (Mukai et al. 2003, Balman \& K\"upc\"u-Yoldas, 2004).
 However, the contribution of the shell to the X-ray flux in
 these novae does not exceed 20\%, so we can reasonably assume that at
 least 80\% of the X-ray emission is due to the point source.

 In the different thermal models in XSPEC
 N(H) does not exceed  2 $\times 10^{21}$
cm$^{-2}$,  the upper limit to the column density to 
 CP Pup  indicated also by Balman et al. (1995).
 The best fit value we obtained
 for the normalization parameter is \mdot= 9.4 $\times 10^{-11}$
 M$_\odot$ yr$^{-1}$,  but it depends on the 
 distance. For d=1600 pc, the 3 $\sigma$ upper limit is \mdot$\leq 1.6
 \times 10^{-10}$ M$_\odot$ yr$^{-1}$, and conservatively we will
 assume this as upper limit to the mass transfer rate
 this nova. However even at the lowest estimated distance, 
 850 pc, the 3  $\sigma$ upper limit is only   \mdot$\leq 8 \times 10^{-11}$
 M$_\odot$ yr$^{-1}$, the same value obtained
 for V603 Aql (Mukai \& Orio 2005). Thus, 
 the possible range of \mdot \ values is thought to be rather on
 the low side for a classical nova. The other interesting result we obtain
 is the high maximum temperature, which is not constrained 
 (80 keV is the maximum value allowed in MCKFLOW), 
 consistently with the spectral fits to the broad band spectra
(see also below).  One important reason
 is that that the  Ne X H-like line at 12.13 \AA \ is much stronger
 than the Ne IX He-like triplet around 13.5 \AA.
 H-like lines are emitted over a wider range of temperature
 than He-like line, and are stronger in a multi-temperature
plasma with a wide range of temperatures.  

 Although values as low
 as 10 keV are within the 2 $\sigma$ error bars because the 
temperature is not well constrained by the RGS spectra,    
 the  high temperature is confirmed by fitting 
 the broad band spectra of the EPIC instruments and
 {\sl ASCA} with either MCKFLOW or 
 a multi-temperature APEC model in XSPEC (see Table 1).
The APEC code in XSPEC calculates the emission spectrum from a collisionally
ionized diffuse gas, using as parameters the plasma
temperature in keV, abundances of several elements (C, N, O, Ne, Mg, Al, Si,
S, Ar, Ca, Fe, Ni), the redshift, and the electron and proton density.  
 More information can be found at http://hea-www.harvard.edu/APEC/.
 The best fit  is obtained with 
a three-temperature APEC model for  the EPIC spectra for the individual
instruments or for all the detectors  together, with or without RGS.
 It is not surprising that APEC fits the broad band spectra better,
 because MKCFLOW does not yield a very meaningful fit where emission
 lines are not resolved. The EPIC spectra on the other hand 
 cover the much larger energy range 0.15-10 keV, so
 the maximum temperature (i.e., the shock temperature) is 
better constrained,
 yet the best fit is still obtained  with a component at
 the maximum allowed value, 64 keV. 
We find that the three {\sl EPIC} spectra together can
 only be fitted assuming a temperature component T$_{\rm max} \geq$50 keV, 
 a 3 $\sigma$ lower limit.
  In a fit to the EPIC broad band spectra with MKCFLOW, not
 included in Table 1 because it is worse than the APEC fit 
 (the reduced $\chi^2$ value is 1.3), 
the minimum temperature is also in agreement with the RGS result,
 and  the \mdot \ values
 are even lower, \mdot=3 $\times$ 10$^{-12}$ M$_\odot$ year$^{-1}$ and
\mdot= 3.7  $\times$
 10$^{-12}$ M$_\odot$ year$^{-1}$ respectively for the EPIC pn
 and MOS data, \mdot= 6.9 $\times$ 10$^{-12}$ M$_\odot$ year$^{-1}$ for
 the {\sl ASCA} data. 
 There is only an upper limit for the
 visual luminosity before the outburst, V$<$17 (Payne Gaposhkin 1964),
 and CP Pup is currently  at much higher optical luminosity
 than before the outburst, as we discuss in Section 3.1, so it seems
 very unlikely that this low \mdot \  value is explained 
with the  ``hibernation'' model (Shara et al. 1986). 
The fact that this nova has not completely returned to 
the quiescent visual luminosity after so many years probably
 indicates that \mdot \ shortly before the eruption
was even less high than now, after 65 years.

 We remark again that {\sl ASCA} flux 
of 1998 (see Table 1) is in agreement with the
 {\sl XMM-Newton} results for all thermal models.
  The {\sl ROSAT} and {\sl Einstein} flux was not well constrained, but it was 
 within the 3$\sigma$ limits  assuming  most models. The flux 
 could not be well determined in those short observations
 with narrow bandpass (especially 
 {\sl ROSAT}) that only sampled the low tail of the X-ray spectrum.
 Variations in N(H) or small changes in the minimum temperature
 of the cooled plasma may also produce a discrepancy.  
 
 We note that the iron K$\alpha$ complex is detected in the EPIC spectrum
(see Fig. 3).
The H-like line and the He-like complex (as a whole) are resolved
 and like for other CV, a fluorescent line at 6.4 keV is
 also present, that is not produced
 in the thermal plasma model.  It originates in ``cold'' iron, in 
 any ionization state, and is due to scattering
 of X-rays (see discussion by Hellier
 \& Mukai, 2004). In the EPIC-pn spectrum this line
 seems to be broad and we note that it shows a hint of a 
large red wing (perhaps extending as much as
 200 eV to the red), suggestive of the ``Compton down-shifted'' shoulder 
like the one detected in GK Per (Hellier \& Mukai 2004). The H-like line
 appears to be quite less prominent than the He-like complex,
 like for Ne in the RGS range.
 The best fit to the EPIC spectra is obtained with an APEC
 thermal model   with three temperature components. Adding
 a new component does not improve the fit. 
The minimum value obtained for the temperature in the
RGS MKCFLOW fit is 0.8 keV,
 and the coolest temperature component is
 at 0.9 keV in the fits to the EPIC instruments. The intermediate 
component temperature in the APEC model fits to
 the EPIC data, like for the ASCA data, is around 5-6 keV for the different instruments.

The most significant parameters obtained from the spectral fits are \mdot
 \ and the highest temperature is T $\geq$50 keV
for all models fitted to the spectra of all detectors.
 This temperature  represents the best estimate of the shock temperature.
  Following Wu et al. (2003, and references therein), in
 a magnetic CV the shock temperature in keV 
 is related to M$_{\rm WD}$ by a simple relationship:

\begin{equation}
  kT \approx 26 \times (0.5/\mu) ({\rm M}_{\rm WD}/{\rm M}_\odot) (10^9 {\rm cm}/{\rm R}_{\rm WD})
\end{equation} 
 
 where $\mu$ is the mean molecular weight. For $\mu \simeq$0.5, 
 if there is no significant expansion of the accreted envelope
 (that is, the WD radius is not extremely bloated, which
 however would cool the envelope layers and disrupt the continuity of H burning)
 the post-shock temperature exceeds 60 keV only for M$_{\rm WD} >$ 
1.1 M$_\odot$, a lower limit for the WD mass. The 
 lower limit to the WD mass is even {\it higher} if CP Pup is
 not a magnetic accretor. 
 For the case of a disk accretor in which
 half of the shock energy is dissipated in radiation,
 Luna \& Sokoloski (2007, and see references therein) show
that the maximum temperature exceeds 50 keV if M$_{\rm WD} >$ 1.3 M$_\odot$,
 assuming no significant variation in radius during 
 quiescent hydrogen burning. 
   
\section{Timing analysis for CP Pup}

 We do not observe for CP Pup the type of irregular aperiodic variability
 on minutes time scales that was observed for V603 Aql, but
 there is a prominent peak in the Fourier spectrum of
 the EPIC-pn and EPIC-MOS light curves, corresponding to three
 periods that are very close the spectroscopic one:  
 0.0602587 days for the pn, and 0.0616597 days for MOS 1,
 0.0620440 days for MOS 2.  The 1$\sigma$ uncertainties 
 in the determination of these periods are about 7 minutes, so not only
 are these three periods well within each other's statistical errors, 
but they are also in agreement with the spectroscopic period determined
 by B08, 0.0612643 days. We also
 split the EPIC pn light curve in three energy ranges, 0.2-0.8 keV,
0.8-2 keV, and 2-10 keV.  The most prominent peak corresponds
 to periods of 0.0601676 d, 0.0612797 d, and
0.0594426 d, in each of the three bands respectively.
 These three periods  are all consistent within the  statistical
 error in the determination. Fig. 5
 shows the EPIC-pn light curve in the 0.2-10 keV range folded with
the period of B08. 
In Fig. 6 we show instead the light curves of the three energy
 bands indicated above, folded with the best X-ray period obtained
 in the whole wavelength range. It is evident that the largest modulation is 
in the 0.8-2 keV range. 
 Using the ephemeris of B08, T$_0$= JD 2450157.6405 and the above period,
 there seems to be a  small phase
 shift between optical-spectroscopic and X-ray period (see Fig.5).  
 The maximum individual value in the X-ray light curve is measured at spectroscopic
 phase $\phi$=0.1,  however the fit with a sine
 function, with an amplitude modulation of (24.8$\pm$0.1)\% with
 respect to the average value, 
indicates a maximum at phase 0.0180$\pm$0.0006.
 The curve shown in Fig. 5 is a fit with a reduced $\chi^2$=1,
 but it may not be meaningful, since  there is an error of
 0.017 s in the determination of the spectroscopic period, which
 may add up to even 20\% of the period in  $\simeq$10 years.
 The spectroscopic phase 0.0 is defined as the epoch of red-to-blue
crossing of the emission line radial velocity curve.
In the baseline interpretation, the spectroscopic period is the
orbital period and phase 0.0 corresponds to the inferior conjunction
of the secondary.  High inclination non-magnetic CVs have an X-ray
minimum around phase 0.0 (eclipse) or around phase 0.75 (dips).
Orbital modulations of X-ray flux have also been observed in many
IPs at similar phases (e.g., Parker et al. 2006).

We used discrete Fourier transformation techniques for a period search.
 Due to the periods of very high cosmic background that we had
 to exclude (approximately a total 35\% of the exposure time), we also
used the CLEAN algorithm (Roberts et al. 1987)  to eliminate spectral peaks
 originating from the temporal distribution of the data. The CLEANed 
 power spectrum does not show any other significant peaks, so we rule out
 modulations of the X-ray flux on time scales of  tens of minutes, 
characteristic of the rotation periods of many IP's. We conclude
 that we could not 
prove the magnetic nature of CP Pup by detecting the WD
 spin period in X-rays. 

  To better examine the spectral variations during
 the orbital period, we plotted the count rate ratio in the 
0.2-0.8 keV, most effected by absorption,  over the count rate
 in the 2-10 keV band, least effected by absorption, and 
 folded this ratio
with the spectroscopic period. We do find a modulation of
 this softness ratio, but the fit with  sinusoidal functions
 in this  case would not be good (attempts  to do it yield 
 a large $\chi^2$). The curve seems to be asymmetric around the maximum
measured value,
 that falls again at spectroscopic phase $\simeq$0.1. Like for many
 magnetic CV, this asymmetry may be real and due to how 
 accretion occurs. This modulation of the
 softness ratio is consistent with absorption modulating 
 the X-ray flux, but the situation may be rather complex
 with several different emitting regions. 
 We extracted the the RGS spectra only at
X-ray phases 0.10-0.25 and 0.75-1, and then we repeated the exercise for
 X-ray phase 0.25-0.75, to study 
 the ``bright'' and ``faint'' half of the period, and we found
 no  difference in the emission lines strength in the RGS spectra,
 even if the count rate is 
 lower in the faint time.  Thus the  explanation in terms of varying absorption
along the line of sight may be too simplistic.
\clearpage
\begin{table}
\rotate
\begin{center}
\caption{\small Observations of CP Pup: instruments, year of observation,
 energy bandpass, count rate, best fit model, and its parameters, 
 including unabsorbed fluxes in two bands with 2$\sigma$  error
 (3$\sigma$ for ROSAT), assuming
 that the value of the plasma temperature
T, or of T$_{\rm max}$ for the multi-temperature
models, is the maximum allowed in the model.
 The models are thermal bremsstrahlung and Raymond-Smith with one component
 (TB and RS, used by Balman et al. 1995;
 see also Raymond \& Smith 1977 for RS, and Kellogg et al. 1975 for TB), 
APEC with three temperature components
(3T), and MKCFLOW (MKCF).  The double line separates
 the previous observations from the new ones we
 present here.  The {\sl ROSAT} values
 are from Balman et al. (1995), the Einstein flux is 
 a PIMMS conversion of the {\sl ROSAT} model.  The reduced
 $\chi ^2$ is between 1 and 1.1 for {\sl ROSAT} and {\sl ASCA}; 
  $\chi ^2$=1.1 for the fits to a single type of
 instrument (the pn, the 2 MOS and the 2 RGS of {\sl XMM-Newton}); 
 $\chi ^2$=1.2 in the simultaneous fit to  
 all EPIC or all {\sl XMM-Newton} instruments.
 The ``(F)'' after the N(H) value indicates
 that a fixed value was assumed.  The MOS and GIS spectra were always fitted
using both detectors. The parameters that are not given in
 the Table are indicated in the text.}
\begin{tabular}{crrrrrrr}
\tableline\tableline
 Instrument & Year & Bandpass & Count rate     & Model & N(H) (10$^{21}$ & T &
Flux (10$^{-12}$  \\
            &      & (keV)    & (cts s$^{-1}$) &       & cm$^{-2})$            & (keV)         &
erg s$^{-1}$) \\
\tableline
  {\sl Einstein} IPC        & 1980 & 0.2-4 & 0.060$\pm$0.006 & TB & 1(F) & 4   & 3.0  \\
  {\sl ROSAT} PSPC  & 1990 & 0.2-2.4 & 0.061$\pm$0.021 &  &   &        & \\
  (survey)          &      &         &                 &  &   &        & \\ 
  {\sl ROSAT} PSPC  & 1992 & 0.1-2.4 & 0.067$\pm$0.004 & TB & $<$2  & 3 & 3.0$^{+8}_{-2}$ \\
                    &      & 0.1-2.4  &                & RS & $<$2  & 1 & 0.9$^{+0.3}_{-0.4}$ \\
\tableline
\tableline
  {\sl ASCA} GIS-2          & 1998 & 0.7-10 &  0.0203$\pm$0.0009   & RS & 1(F) & 64
  & 6.1$^{+0.3}_{-1.2}$ \\
                            &      & 0.7-2.4 &                     & RS & 1(F) & 64 & 2.0 \\
  {\sl ASCA} GIS-3          & 1998 & 0.7-10 &  0.0412$\pm$0.0012   & RS & 1(F) & 64  & 6.1$^{+0.3}_{-1.2}$
 \\
 {\sl ASCA} GIS-2/3         & 1998 & 0.7-10 &                      & TB & 1(F) & 199 & 6.4$\pm$1.2 \\
                            &      & 0.7-2.4 &                     & TB & 1(F) & 199 & 2.0 \\
 {\sl ASCA} GIS-2/3         & 1998 & 0.7-10 &                      & 3T & 2.3 &  64 & 6.7$\pm$2.0 \\
                            &      & 0.7-2.4&                      & 3T & 2.3 &  64  & 2.0$\pm$1.9 \\    RGS-1 & 2005 & 0.33-2.5 & 0.0195$\pm$0.0015 & MKCF & 1.92 &  78  & 2.3$\pm$0.2 \\ 
 RGS-2 & 2005 &  0.33-2.5 & 0.0249$\pm$0.0016 & MKCF & 1.92  & 78  & 2.3$\pm$0.2 \\
 RGS-1/2 &   2005  & 0.33-2.5 &                  & 3T & 1.6(F) & 77 & 1.4$\pm$0.2 \\
 EPIC-pn & 2005 & 0.15-10 & 1.3090$\pm$0.0074 & 3T & 1.43 & 64 & 6.8$\pm$0.7  \\
         &      & 0.15-2.4 &                  & 3T & 1.43 & 64 & 2.4$\pm$0.1  \\ 
 MOS-1   & 2005 & 0.3-10 & 0.3782$\pm$0.0036 & 3T & 1.73 &  64 & 7.3$\pm$0.8 \\   
 MOS-2 & 2005 & 0.3-10 & 0.4251$\pm$0.0035 & 3T & 1.73 &  64 & 7.3$\pm$0.8 \\
 MOS  & 2005  & 0.3-2.4 &                   & 3T & 1.73 & 64  & 2.4$\pm$0.2 \\
All EPIC   &   2005 & 0.3-10 &    &  3T & 1.53 &  64 & 6.9$\pm$0.7  \\
           &        & 0.3-2.4 &   & 3T  & 1.53 &  64 & 2.4$\pm$0.3  \\
All XMM    &  2005  & 0.33-2.5 &   &  3T & 1.34 & 64 & 2.2$\pm$0.3 \\
\tableline
\end{tabular}
\end{center}
\end{table}
\begin{table}
\begin{center}
\caption{Magnitudes measured for CP Pup
with the {\sl XMM-Newton} OM on June 4 and 5, 2005
 (using the OMICHAIN XMM-SAS task). We indicate also the  
 spectroscopic phase $\phi$ at the
 beginning of the observation and the portion of phase $\Delta\phi$ spanned 
 during the exposure.}
\begin{tabular}{crrrrrr}
\tableline\tableline
 Filter & Initial time & JD & Exposure (s) & Magnitude & $\phi$ & $\Delta\phi$ \\
\tableline
 U    & 15:09:50 & 2453526.1318287 & 1008 &  14.133$\pm$0.005 & 0.94 & 0.19 \\
 U    & 16:14:59 & 2453526.1770718 &  1000 & 14.090$\pm$0.005 & 0.68 & 0.19 \\
 U    & 17:20:08 & 2453526.2223148 &  1001 & 14.084$\pm$0.005 & 0.42 & 0.19 \\
 B    & 18:25:17 & 2453526.2675579 & 1001 & 15.414$\pm$0.007 & 0.16 & 0.19 \\
 B    & 19:30:27 & 2453526.3128125 & 1001 & 15.389$\pm$0.007 & 0.89 & 0.19 \\
 UVW1 & 21:35:35 & 2453526.3997106 &  2001 & 14.154$\pm$0.006 & 0.31 & 0.38 \\
 UVW1 & 22:27:25 & 2453526.4357060 & 4999 & 14.123$\pm$0.004 & 0.90 & 0.94 \\
 UVW1 & 1:09:13 &  2453526.5480671 & 2001 & 14.137$\pm$0.006 & 0.60 & 0.38 \\
 UVM2 & 2:31:03 &  2453526.6048958 & 3124 & 14.602$\pm$0.013 & 0.66 & 0.59 \\
 UVM2 & 4:11:37 &  2453526.6747338 & 2000 & 14.572$\pm$0.016 & 0.67 & 0.38 \\
\tableline
\end{tabular}
\end{center}
\end{table}
\clearpage
\subsection{CP Pup: Optical Monitor measurements and comparison with IUE}
  Photometry was done during the X-ray exposures with
 the Optical Monitor and the B, U, UVW1 and UVM2 filters.
 The exposures were all longer than 1000 seconds, which is a significant
 fraction of the $\simeq$5290 s spectroscopic
and putative orbital period, so no variability
 was detected in any of the filters in two or three successive
 exposures. The results are reported in Table 2. 
 In the optical range, 
there are few terms of comparison, but we note that the USNO
 magnitudes in 1978 were B=14.1 and R=13.7,
 and  Fig. 3 of White et al. (1993) 
 shows that in 1987, CP Pup had an average value of  B$\simeq$14.8,
 with a total amplitude of the photometric
 modulation of about 0.43 magnitudes.
The magnitude B=15.4 in Table 2 is significantly higher,  
and implies that CP Pup is slowly returning to pre-outburst
 luminosity.
Photometric monitoring is strongly encouraged to
understand the long term evolution of this nova.

 While the B magnitude shows a significant decrease, the UV flux
 seems to have remained approximately constant in recent years.
 CP Pup was observed with IUE.  Exposures with the SWP
 (Short Wavelength Prime Camera) and with the LWP (Long
 Wavelength Prime Camera) were taken 
 on 1986 February 26 and a SWP exposure was done on 1992 May 27.
 In order to compare the flux, we have to refer to the measurements
 with the LWP, which partially overlaps with the OM filters.
The LWP flux in the range
 of the UVW1 filter is (1.01$\pm 0.3) \times 10^{-14}$
 erg cm$^2$ s$^{-1}$ \AA$^{-1}$ 
(averaged on  an exposure of 5100 s, a time comparable
 with the spectroscopic period).
The OM magnitudes in Table 2 for the 
UVW1 OM filter (2450-3200 \AA, effective wavelength=2910 \AA) 
correspond to a flux 8.50 $\times 10^{-15}$ \AA$^{-1}$ 
 erg cm$^2$ s$^{-1}$ \AA$^{-1}$ and the UVM2 
 flux (2000-2600 \AA, effective wavelength=2310 \AA) is 6.45  $\times 10^{-15}$ 
 erg cm$^2$ s$^{-1}$ \AA$^{-1}$ . 
The OM exposures were  about 1000 s long, covering a significant
 portion of the  spectroscopic phase.
 We conclude that there is no evidence of flux change in
 the common wavelength range covered 
by IUE-LWP and by  the UV filters of the OM between 1986 and 2005.
  We will make a few other comments about the IUE archival spectra,
 that we examined in detail. 
Matching SWP and LWP spectra of 1986  to evaluate the absorption
 is not possible because the  data
are very noisy in the 2200 \AA \ region.
The two SWP exposures
 of different epochs are comparable
 in length (18000 s in 1986 and 18900 s
in 1992) but in 1992 the SWP spectrum showed approximately a
decrease by 30\%, so there
 may be variability on long time scales. 
 Although the SWP spectra had poor S/N, 
 emission lines of  N V, O I, C III, Si IV, C IV, He II, N IV 
and possibly O V  at 1372 \AA \ were  definitely detected. 
Other IP, and all magnetic CV in general,
 show the same  emission lines (Howell et al. 1999), 
 but a reliable characterization of the spectra,
 with the poor S/N obtained, is not possible.

\section{Discussion of the CP Pup results}

 We find that the X-ray spectrum of CP Pup indicates 
 a very high maximum temperature (unbound by the models), 
consistent with a very high mass WD.
Equation (1) yields M$_{\rm WD}>1.1$ M$_\odot$ for an IP. 
 Under ``standard'' assumptions for accretion only through a disk,
 we derive M$_{\rm WD}>1.3$ M$_\odot$. We also find that
the WD must be accreting at relatively low rate
 compared to the estimates for other novae, \mdot$<1.6 \times
 10^{-10}$ M$_\odot$ yr$^{-1}$. Since we do not observe a hot central
 source, the secondary is not irradiated and it is reasonable
 to assume that \mdot \ is still, or
 has returned to, the pre-outburst level. The
 value we derive for \mdot \ is in agreement 
 with  the models of Prialnik \& Kovetz (1994) and Kovetz
\& Prialnik (1995) for a fast nova on a high mass WD. 
The nova parameter space was explored by this group in the two
 papers above and in Yaron et al. (2005). According to these authors 
 t$_2$ and t$_3$ are inversely proportional to the WD mass, and a large 
 amplitude nova with moderately high ejection velocity 
 occurs on a massive
 WD only if \mdot$\leq 10^{-10}$  M$_\odot$ yr$^{-1}$,
 which is believed to be lower than \mdot \ of most novae. 
CP Pup fits this predictions very well.
 In Epelstein et al. (2007) the authors make another prediction,
 namely that a massive WD becomes increasingly much hotter with the number 
of nova outbursts it experiences. A luminous and fast eruption 
 is only possible at the beginning of the outbursts' cycle. If
 this is true, CP Pup must be at the beginning
 of the nova cycle, and the next outburst (which however will occur
 in more than 10$^5$ years due to the low \mdot) will already be less
luminous.  
 In any case, the  large WD mass coupled with low \mdot \ probably
explains why the ejecta velocity was not as high as in some other
 large amplitude, luminous novae. In this respect our X-ray observations
 nicely confirm the nova theory.

 We measured an X-ray flux modulation of about
 50\% amplitude with the optical spectroscopic period,
generally thought to be the orbital period . We note
 that also the optical and IR modulations have
 approximately the same amplitude. This similar behavior 
in  very different energy bands cannot be easily
explained with the superhump model, neither with ellipsoidal
 variations.

If the inclination of CP Puppis is less than
 35$^{\rm o}$, as inferred with the ellipsoidal 
 variation interpretation of the light curve
 (see Section 2), no  orbital modulations are expected for a 
disk accretor. Thus, our baseline interpretation, that the 1.47 hour
 spectroscopic
period is the orbital period, encounters a series of difficulties.
If CP Pup is non-magnetic, then the emission line radial velocity
curves suggest an unrealistically low-mass white dwarf, and we have
no explanation for the X-ray "orbital" modulation.  If CP Pup is
an IP, the X-ray "orbital" modulation is less of a problem, since
X-ray orbital modulations are much more common in IPs, with multiple
origins being discussed (Parker et al. 2006).  However, in this case,
the complete lack of the spin period (at P$<$1.47 hr) is a severe
problem.

We therefore propose an alternative scenario, in which the 1.47 hour
spectroscopic period is the spin period of the magnetic white dwarf, and the
true orbital period has never been detected in the observations to date.  
 In this case,
 we have indeed detected the WD spin in X-rays, and more optical
 observations are needed instead to discover the true orbital period of CP
 Pup.  Without invoking superhumps,
the difference in spectroscopic and photometric periods
 of CP Pup is naturally explained if the modulations of the flux and
of  the spectral lines are caused by variations in the accretion
 disk, as it is periodically illuminated by the hot polar caps of an IP
  while the WD rotates. One possible model is the one by Chanan et al. (1978)
proposed by Penning (1985) to explain the radial velocity periods of
 four different IPs. Another example of spectral line variations in
 a rotating WD  is the EX Hya, another IP (Belle et al. 2005).
The optical emission lines in these systems vary periodically
because of the illumination effect, so that the phasing of
 the X-ray flux with the optical spectroscopic period is quite close.
 This interpretation avoids the mass problem for CP Pup, and
explains why the attempt to determine the mass of the WD
from the mass function, assuming that the spectroscopic period is the 
orbital one, gives an unreasonable result. 

 The question is of course why the true orbital period has
 remained undetected and in which range it should be searched.
The system has a sizable disk, because the optical
 emission lines often show the saddled profile. In 
 an IP, the disk is disrupted at the
 Alfv\`en radius. If the spin period is about an hour and a half,
 the orbital period is expected to be at least $\approx$10 times larger.
 We also note that, if the photometric period is the beat of the orbital
 and spin period, the orbital period turns out to be 16.17 hours. 
However, CP Pup has an undetected secondary. With an orbital period
 larger than 10 hours, the secondary is evolved and is always detected
 in CV at d$\leq$2 kpc. The non-detection of the secondary in
 the optical spectrum therefore puts another serious constraint on the length of
 the orbital period. It seems difficult to admit that it is
 sufficiently longer than a night-time span to have not been detected yet
 in 3-4 nights runs.  

 Could the orbital period instead
 be short and yet have gone undetected in optical?
 It seems that this may be true only if CP Pup is an EX Hya-like system,
 that is an IP with spin period that
 is shorter by only less than 40\% than the orbital one
 (Norton et al. 2004, 2008). 
 Instead of a real accretion disk, in these systems there is 
 a ``ring-like'' structure, rotating at non-Keplerian velocity. 
 The possibility that disk of CP Pup is a thin ring of this kind
is indeed intriguing. 
 The profile of the optical emission lines of CP
 Pup is not very definite and simple
 to interpret (B08). The emission lines of the proto-type
 system EX Hya (Belle et  al. 2005) appear to have a saddled profile
 like expected in ``traditional'' disks, and observed for
 the optical emission lines in the
 CP Pup optical spectrum. 
 In a EX Hya type system, orbital and spin
 period can be even very close in length. This
 may be the case of CP Pup. For the time being, this is
 only a speculative suggestion, but it certainly deserves renewed optical
 monitoring to be explored. 

\section{Nova Puppis 1991}

The {\sl ASCA} upper limit
to the X-ray flux of this nova is 10$^{-13}$ erg cm$^{-2}$ s$^{-1}$
in the 0.3-10 keV range, much below the value 
 F$_{\rm x}=3 \times 10^{-12}$ erg
cm$^{-2}$ s$^{-1}$ observed with {\sl ROSAT} in 1993 April.
We detected V351 Pup in 2005 serendipitously with {\sl XMM-Newton,}
 at the edge of the   EPIC detectors.
The MOS1 count rate is 0.0015$\pm$0.0004 cts s$^{-1}$.
 The source is even more peripheral in the pn observation,
 and is not fully observed, so the count rate is only 0.0032$\pm$0.0010 cts s$^{-1}$.
The flux measured with the MOS is 6 $\times 10^{-14}$ erg cm$^{-2}$ s$^{-1}$
 in the 0.2-10 keV range. 
Assuming a Raymond-Smith model of thermal plasma emission
with the value of the column density 2 $\times$ 10$^{21}$ cm$^{-2}$ 
measured by Orio et al. (1996), the unabsorbed flux 
 is F$_{\rm x} = 7 \times 10^{-14}$ erg cm$^{-2}$ s$^{-1}$. 
 a factor of 50 less than in 1993 April. 
 The X-ray luminosity does not exceed 3 $\times$ 10$^{31}$  
erg s$^{-1}$ in the {\sl XMM-Newton} observation.
The most likely interpretation is 
that the X-ray flux was due to the ejecta back in 1993, which by 2005 
 had cooled. The X-ray emission observed in 2005 is most likely
 due to accretion, although the poor quality of the data does not
 allow any definite conclusion.
 
\section{Conclusions}
 CP Puppis is only the second classical nova for which
 an X-ray grating spectrum could be obtained at quiescence. Even if with 
 low S/N, this spectrum shows strong emission lines which
 probably mainly arise in a cooling flow, either in the thermal
plasma of disk or of accretion columns of a magnetic 
 WD. There is at least one emission complex
 (O VII at 21.6-21.8 \AA), that most likely arises 
 instead in a wind from the system, or in a circumstellar nebula. 
CP Pup had a very large outburst amplitude, a short t$_3$, and
 only moderately high ejection velocity. 
 Examining a number of theoretical
 papers quoted above, we find that these characteristics are
 explained by the upper limit to the mass accretion
 rate \mdot$< 1.6 \times 10^{-10}$
M$_\odot$ year$^{-1}$, and by the lower
 limit for the WD
 mass, M(WD)$>$1.1 M$_\odot$, obtained by fitting the
 X-ray spectra.  The X-ray observation of this nova at
 quiescence therefore confirm the nova theory, offering
 a rare possibility to probe it in detail. 
 The models of WD thermonuclear burning and runaways are also at the base 
of type Ia SN studies, and testing them is important for
 all aspects of modern astrophysics, including cosmology. 

   The puzzle of the WD mass of CP Puppis is not solved yet.
 With our observations, we could not demonstrate 
 univocally that the WD is magnetic and the disk is disrupted far
 from the WD, however   the high L$_{\rm x}$ is typical for an IP.
The only point we found against
 the magnetic nature is relevant only if the detected X-ray modulation
 is orbital: the X-ray orbital modulation of CP Pup is larger
at intermediate energy (0.8-2 keV), 
 although
in known  IP's the amplitude of orbital modulations generally decreases
 with energy (e.g. Hellier et al. 1993, Parker et al. 2006).
 On the other hand, the mere existence of an X-ray flux  
 modulation with the optical-spectroscopic period
argues in favor of an IP, no matter how it is interpreted. If it is orbital,  
 it would not be detected in  a non-eclipsing disk accretor.
 If it is rotational, detection of the WD spin is indeed expected for an
IP.  The high WD mass is also proof that the non-magnetic, disk
 accretor model does not work for CP Pup. 
 It cannot be
 ruled out that all the observed periodicities 
at different wavelengths, including our X-ray
modulation, are related to the WD rotation and the true orbital 
 period has never been detected. 
   This suggestion, for the time being,
 is only a working hypothesis that will need to be supported by further
 observations, especially by long optical photometric campaigns.  
   
 Finally, we also detected V351 Pup 14 years after the outburst, 
 at least 50 times less luminous than 12 years earlier. This seems
 to indicate that the much larger hard X-ray luminosity at that time was due
 to the ejected shell, which had already
 cooled in 2005. Nova shells are known to be conspicuous
 X-ray sources after the outburst  (see for instance the shell
 emission of  RS Oph, Nelson et al. 2008) and in most novae
they seem to cool very rapidly (Orio et al. 2001).
 An upper limit for the cooling time
 of the V351 Pup shell is 6 and a half years, the time elapsed
 between the {\sl ROSAT} and {\sl ASCA} observation. 
\acknowledgments

{\it Facilities:} \facility{XMM-Newton, ASCA.}

\clearpage

%% the file, uncomment the \plotone, \plottwo, and
%% If you need a layout that cannot be achieved with \plotone or
%% \plottwo, you can invoke the graphicx package directly with the
%% \includegraphics command or use \plotfiddle. For more information,
%% please see the tutorial on "Using Electronic Art with AASTeX" in the
%% documentation section at the AASTeX Web site,
%% http://www.journals.uchicago.edu/AAS/AASTeX.
%%
%% The examples below also include sample markup for submission of
%% supplemental electronic materials. As always, be sure to check
%% the instructions to authors for the journal you are submitting to
%% for specific submissions guidelines as they vary from

%% This example uses \plotone to include an EPS file scaled to
%% 80% of its natural size with \epsscale. Its caption
%% has been written to indicate that additional figure parts will be
%% available in the electronic journal.

\begin{figure}
\begin{center}
\includegraphics[width=10.5cm,angle=-90]{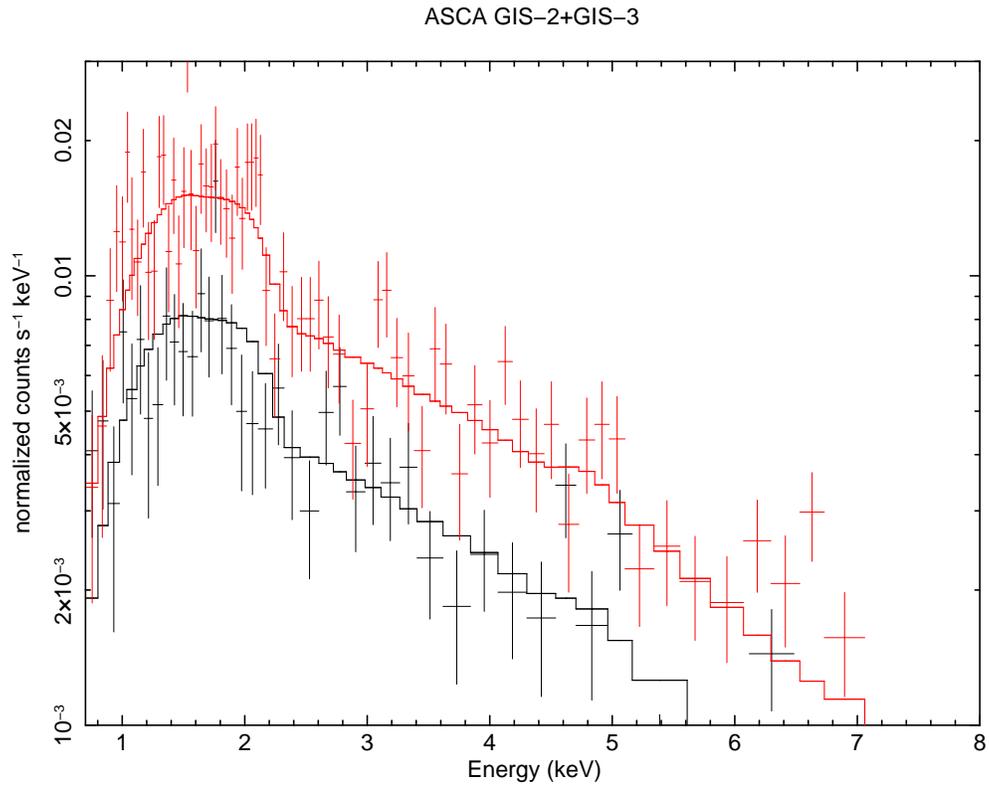}
\end{center}
\caption{The spectrum of CP Pup observed 9 arcmin
off-axis with the two {\sl ASCA} GIS (GIS-2 in black and GIS-3 in red)
 in 1998 November,
 and a fit with  the Raymond-Smith model described in the
text and in Table 1.} 
\end{figure}
\clearpage
\begin{figure}
\begin{center}
\includegraphics[width=10.5cm,angle=-90]{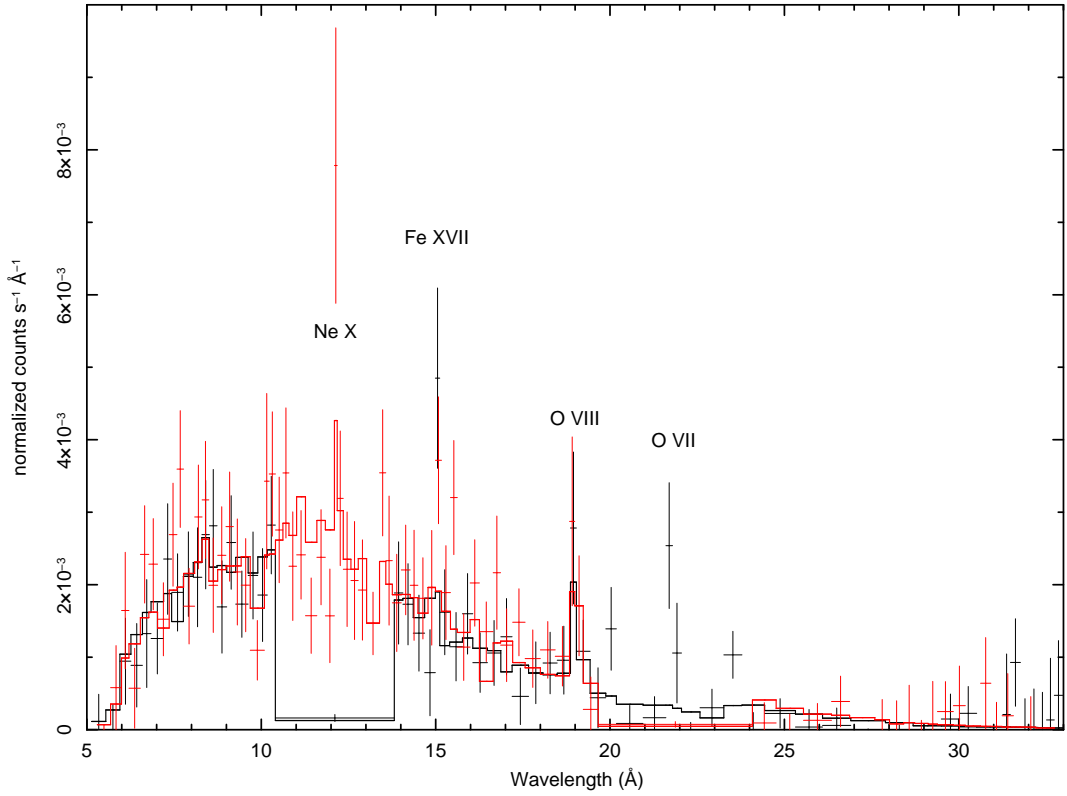}
\end{center}
\caption{The spectrum of CP Pup observed with the RGS-1 (in black) and
with the RGS-2 (in red) gratings of {\sl XMM-Newton} in 2005 June,
 and the fit with a cooling
flow model (see Table 1).} 
\end{figure}
\clearpage
\begin{figure}
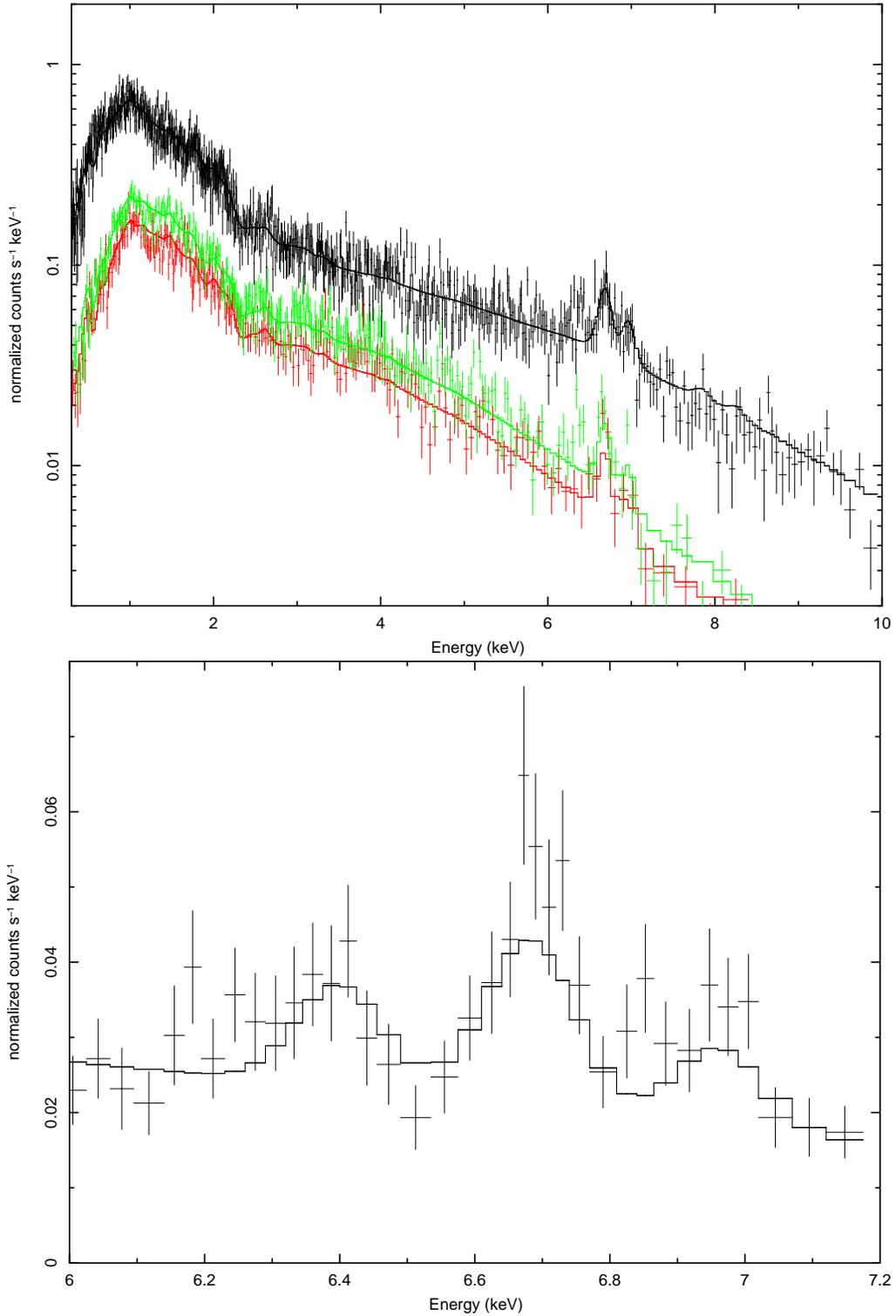

\begin{center}
\includegraphics[width=10cm,angle=-90]{f3a.ps}
\hskip 1cm
\includegraphics[width=10cm,angle=-90]{f3b.ps}
\end{center}
\caption{The spectrum of CP Pup observed in June of 2005
with EPIC-pn (in black) and with MOS-1
 and MOS-2  (in black and green respectively). The
 best fit with the fit APEC model described in the text and in Table 1 is
 also shown. The lower panel 
 shows a close-up of the EPIC-pn iron line region and a fit with
  a slightly different
 three-temperature APEC model and a Gaussian with a 200 eV width for
 the 6.4 keV line.}
\end{figure}
\clearpage
\begin{figure}
\begin{center}
\includegraphics[width=12cm]{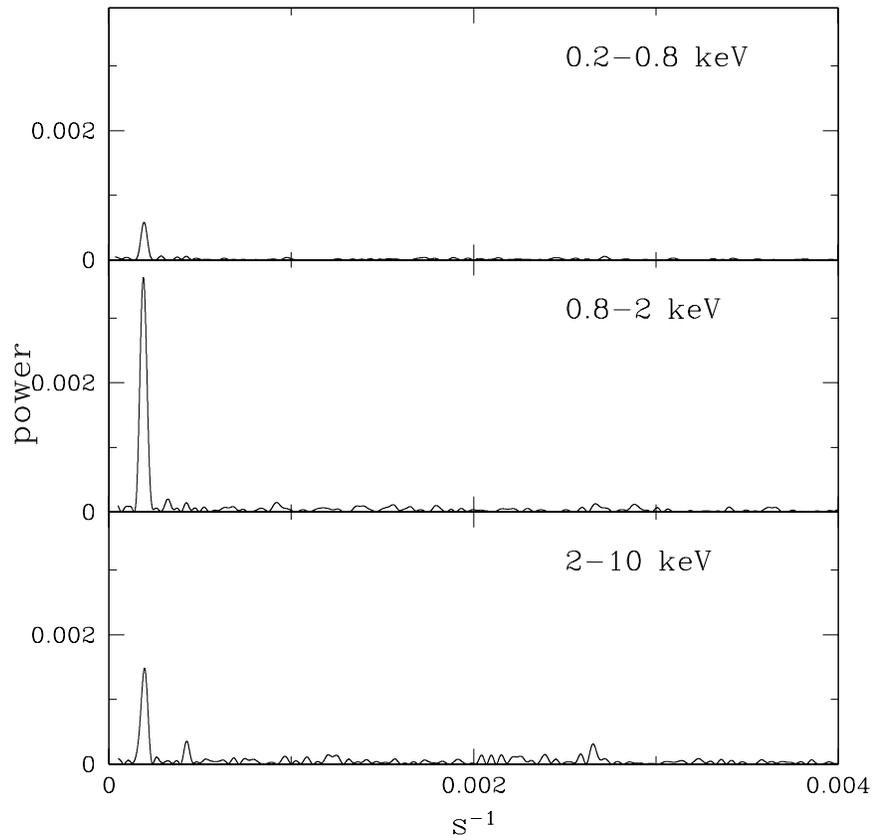}
\end{center}
\caption{The EPIC-pn power spectrum of the EPIC-pn light curve of CP Pup
 in three different energy ranges: 0.2-0.8 keV, 0.8-2 keV,
 and 2-10 keV. The most prominent peak corresponds to the 
 spectroscopic period.}
\end{figure}
\clearpage
\begin{figure}
\begin{center}
\includegraphics[width=12cm]{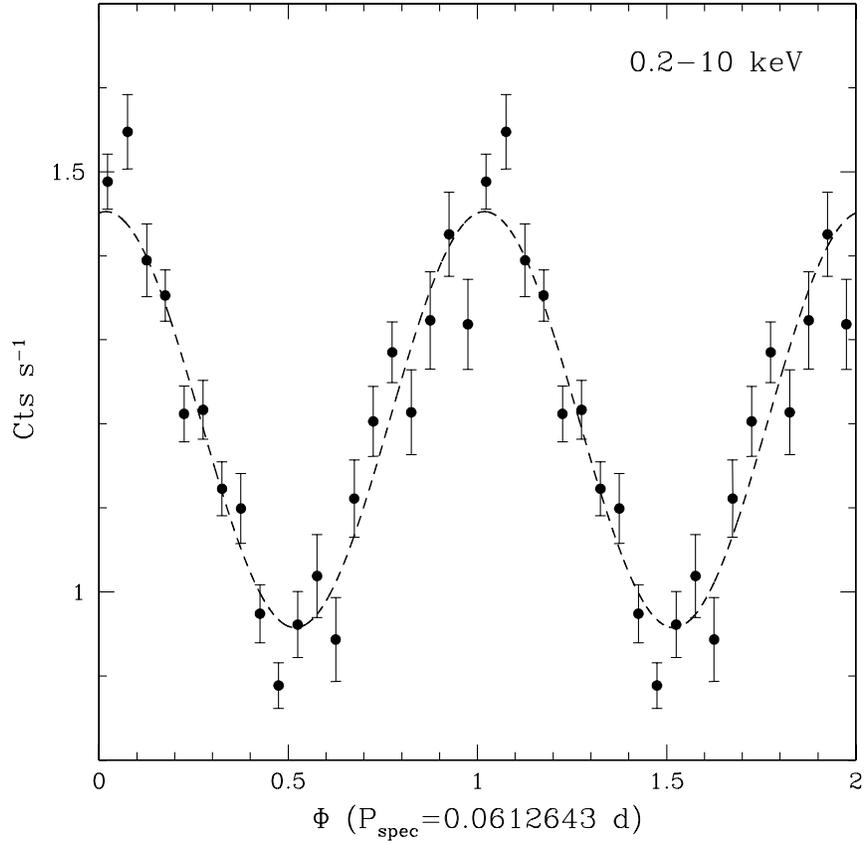}
\end{center}
\caption{The EPIC-pn light curve of CP Pup in the 0.2-10 keV energy range,
folded with the spectroscopic period measured by B08 (see text).
 Using the ephemeris of B08,
 we how a fit with a sine function that has a maximum at phase 0.018$\pm$0.006.}.
\end{figure}
\clearpage
\begin{figure}
\begin{center}
\includegraphics[width=12cm]{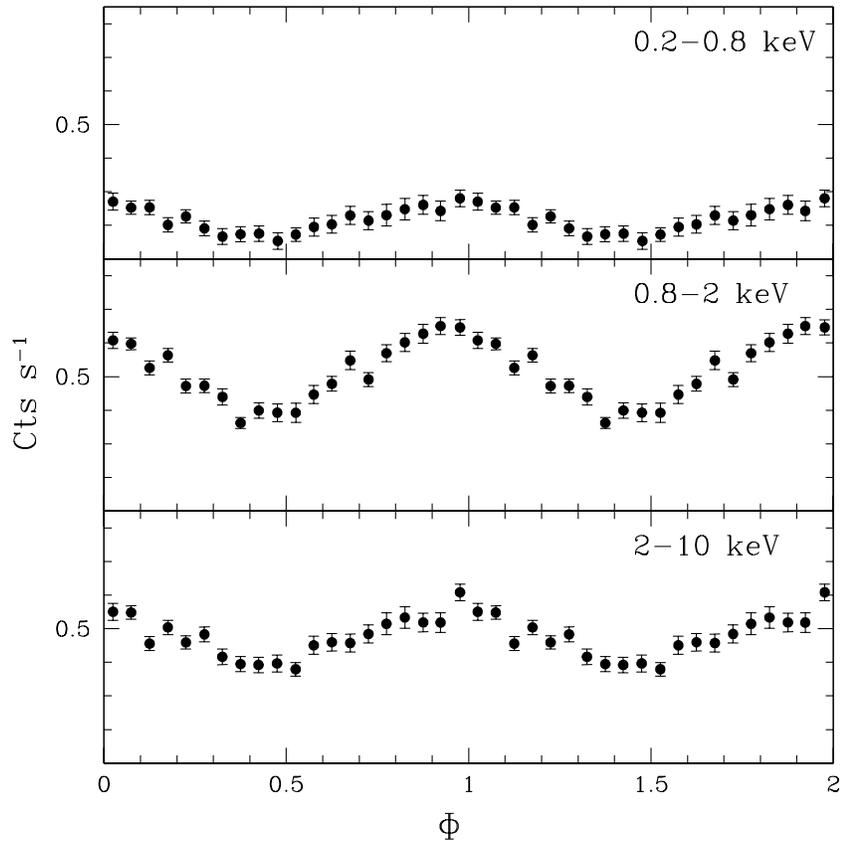}
\end{center}
\caption{The EPIC-pn light curves of CP Pup
 in the three energy ranges: 0.2-0.8 keV,
0.8-2 keV, 2-10 keV respectively, folded with the period found 
 in the Fourier transform of the EPIC-pn light curve (which is very close
 to the spectroscopic one.)}
\end{figure}
\clearpage
%
%\begin{figure}
%\begin{center}
%\includegraphics[width=10.5cm,angle=-90]{f7.ps}
%\end{center}
%\caption{The iron lines region observed at phases 0-0.25 and 0.75-1 (in
% red) and at phase 0.25-0.75 (in black).}
%\end{figure}
%\clearpage
%
\begin{figure}
\begin{center}
\includegraphics[width=10.5cm]{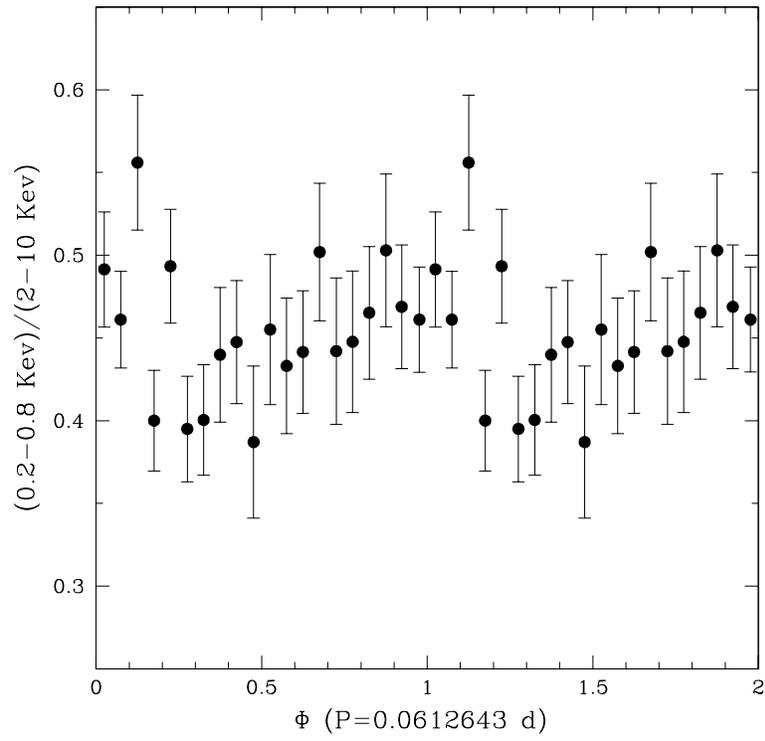}
\end{center}
\caption{The ratio of CP Pup countrates in the 0.2-0.8 keV range and 
 in the 2-8 keV range, folded with the spectroscopic period.}
\end{figure}
\clearpage
\end{document}